# Real-time detection of Ochratoxin A in wine through insight of aptamer conformation in conjunction with graphene field-effect transistor

Nikita Nekrasov[a#], Stefan Jaric[b#*], Dmitry Kireev[c], Aleksei V. Emelianov[a], Alexey V. Orlov[d], Ivana Gadjanski[b], Petr I. Nikitin[d], Deji Akinwande[c] and Ivan Bobrinetskiy[a,b*]

[a]National Research University of Electronic Technology, Moscow, Zelenograd, 124498, Russia, 8141147@gmail.com

[b]BioSense Institute - Research and Development Institute for Information Technologies in Biosystems, University of Novi Sad, Novi Sad, 21000, Serbia

[c]Department of Electrical and Computer Engineering, The University of Texas at Austin, Austin, TX, USA

[d]Prokhorov General Physics Institute of the Russian Academy of Sciences, 119991, Moscow, Russia; petr.nikitin@nsc.gpi.ru

*Corresponding authors: e-*mail*: bobrinet@gmail.com; e-*mail*: sjaric@biosense.rs

# these authors contributed equally

**Abstract**

Mycotoxins comprise a frequent type of toxins present in food and feed. The problem of mycotoxin contamination has been recently aggravated due to the increased complexity of the farm-to-fork chains, resulting in negative effects on human and animal health and, consequently, economics. The easy-to-use, on-site, on-demand, and rapid monitoring of mycotoxins in food/feed is highly desired. In this work, we report on an advanced mycotoxin biosensor based on an array of graphene field-effect transistors integrated on a single silicon chip. A specifically designed aptamer against Ochratoxin A (OTA) was used as a recognition element, where it was covalently attached to graphene surface via pyrenebutanoic acid, succinimidyl ester (PBASE) chemistry. Namely, an electric field stimulation was used to promote more efficient π-π stacking of PBASE to graphene. The specific G-rich aptamer strand suggest its π-π stacking on graphene in free-standing regime and reconfiguration in G-quadruplex during binding an OTA molecule. This realistic behavior of the aptamer is sensitive to the ionic strength of the analyte solution, demonstrating a 10-fold increase in sensitivity at low ionic strengths. The graphene-aptamer sensors reported here demonstrate fast assay with the lowest detection limit of 1.4 pM for OTA within a response time as low as 10 s, which is more than 30 times faster compared to any other reported aptamer-based methods for mycotoxin detection. The sensors hold comparable performance when operated in real-time within a complex matrix of wine without additional time-consuming pre-treatment.

**Keywords:** graphene, mycotoxin, field-effect transistor, biosensor, aptamer, small molecules, ionic strength

**Introduction**

The monitoring of small molecules (*i.e.*, of the size less than 1 kDa) has gained increasing interest in the last years in various areas of science and technology. Understanding the fundamental mechanisms of binding kinetics in molecular biology, as well as accurate and fast diagnostics of contamination in environmental, health, agriculture, and food control, demand for new methods of small molecules detection. Among the small molecules, those of high interest are drugs, antibiotics, pesticides, and toxins. Mycotoxins (MT) are of the utmost interest for health and agriculture monitoring. Mycotoxins are secondary metabolites produced by microfungi such as *Aspergillus* and *Penicillium genera* and one of the most common and dangerous contaminants in food, feed, and agricultural products (Guo et al., 2020; Mata et al., 2015). MT can be found in a variety of foodstuffs, including cereals, dried fruit, and drinks (Sheikh-Zeinoddin and Khalesi, 2019). Ochratoxin A (OTA) is one of the prevailing MT in food and feed. The standard laboratory methods like high-performance liquid chromatography-mass spectrometry (HPLC-MS) and enzyme-linked immunosorbent assay (ELISA) are traditionally used for mycotoxin detection (Zheng et al., 2006). Despite their high accuracy and sensitivity, these methods are technologically demanding in the sense of the equipment and operator qualification, which hinders their in-field application (Dong et al., 2020). Considering the widespread of mycotoxins and their effect on economics, a rapid and low-cost analytical tool for the on-site detection of mycotoxins is highly demanded (Guo et al., 2020).

Graphene field-effect transistors (GFETs) have been demonstrated as a promising platform for the rapid or real-time detection of small molecules (Béraud et al., 2021; Bobrinetskiy and Knezevic, 2018). One of the specific features in the use of graphene sensor for small molecule detection is the large intrinsic charge of molecules that provide reasonably high shift in the graphene channel's conductivity. GFETs have previously been used for the highly sensitive detection of various antibiotics (Chen et al., 2019), small antigens (Kanai et al., 2020), and small-molecule drugs (Xu et al., 2021). While sensitivity and response times are mainly defined by the physical nature of the graphene channel, selectivity is provided by the specific recognition elements, *i.e.*, bioreceptors such as enzymes, antibodies, and aptamers. Aptamers are the most prospective bioreceptors that can be selected through a highly scalable process of systematic evolution of ligands by exponential enrichment (SELEX) (Dunn et al., 2017). Aptamers are oligonucleotides in a typical size range of 20 to 70 nucleic bases that form a specific three-dimensional structure during analyte capture. The oligonucleotides have intrinsic charges distributed over their backbone, and their deformation can bring charges closer to or further from the graphene surface, providing the sensing signal (Chen et al., 2019; Kanai et al., 2020). Different sensing principles based on aptamers have been demonstrated for mycotoxin detection, such as fluorescence (Liu et al., 2018), surface plasmon resonance (Zhu et al., 2015), spectral-phase interferometry (Nekrasov et al., 2021), chemiluminescence (Wang et al., 2019), and electrochemistry (Zhu et al., 2018). To the best of our knowledge, the investigation of environmental effect on the sensitivity of aptamer-linked GFETs to MT has not been sufficiently addressed up to now. Recently, we have demonstrated a proof of concept of GFET with a liquid gate as a sensor for MT (Nekrasov et al., 2019), showing a detection limit of 10 pM and signal detection time of ~5 min. However, the reproducibility of the data is affected by graphene functionalization and aptamer attachment, greatly affecting the electrical properties among different transistors. The use of GFETs with a large area graphene channel (Fakih et al., 2020) could potentially solve this problem, yet it would substantially decrease the overall sensitivity of the device.

In this work, we report on an array of optimized GFET-based aptasensors on a single chip used to detect mycotoxins that can provide better statistical data analysis generated by the sensors. Since the size of an aptamer is larger than the size of a small molecule, such as MT, the main influence on charge carriers in graphene is the transformation of the aptamer's 3D configuration due to the binding of small molecules (Chen et al., 2019). We verified this assumption by demonstrating that buffer's ionic strength dramatically affects the overall sensitivity of the reported GFET aptasensors. We show that diluting the buffer can increase the overall device sensitivity by order of magnitude. Such advanced technology allowed us to track the processes of mycotoxin binding in real-time. The novel graphene-aptamer sensors demonstrate rapid (about 10 s) and superior sensitivity to OTA with the limit of detection (LOD) of 1.4 pM. The cross-selectivity was measured with aflatoxin M1 (AFM1), and almost no response for a similar concentration was found. Furthermore, the devices were tested with real samples of wine, which is one of the major sources of OTA dietary intake in EU region (Zhu et al., 2015), showing inspiring results for commercial application.

## 2. Material and methods

*2.1 Materials and reagents*

The monolayer graphene on copper foil (25 μm thick) was purchased from Grolltex (USA). Solutions of 10 μg/mL of OTA in acetonitrile and 0.5 μg/mL of AFM1 in acetonitrile were purchased from Sigma-Aldrich (USA). Anti-OTA aptamer with sequence GAT CGG GTG TGG GTG GCG TAA AGG GAG CAT CGG ACA (Cruz-Aguado and Penner, 2008) with amino-modified 5' end and purified by HPLC was purchased from Metabion AG (Germany) and Evrogen (Russia). 1-Pyrenebutyric acid N-hydroxysuccinimide ester (PBASE) was acquired from Lumiprobe RUS Ltd (Russia). Ethanolamine and tablets of phosphate-buffered saline (PBS) were acquired from Sigma-Aldrich (USA). Dimethylformamide (DMF) and isopropyl alcohol (IPA) were purchased from Component-Reactive (Russia). Dry red wine and semi-sweet white wine were bought in a local store.

*2.2 Fabrication of the GFETs array chips*

The 4-inch silicon wafers were used to produce chips (52 chips per wafer), each containing an array of GFETs using the high-throughput transfer technique described in (Emelianov et al., 2018; Kireev et al., 2016). In brief, the single-layer graphene was transferred onto a Si substrate with 300 nm $SiO_2$ layer by a wet transfer and then patterned to form graphene channels via oxygen plasma etching (300 W, 200 sccm, 10 min). Using e-beam-assisted evaporation of metals and lift-off of LOR-3B and AZ-5209-E photoresists, we deposited the 10 nm Ni and 70 nm Au metal stack. At the final step, a photostructurable resist SU-8 2002 (MicroChem) was spin-coated to form a ~2 μm thick passivation layer. After exposure, development, and post-exposure baking, the passivation layer, covering the metal feedlines as well as a partial area (<2 μm) of graphene–metal contacts, was formed to prevent current leakage during measurements in a liquid.

To perform multiplex measurements, the chips were wire-bonded (K&S 4524 Wire bonder) on a printed circuit board (PCB). To prevent solvent leakage during measurement and save the reagents during the biosensor assembly, a 5 mm diameter well punched in PDMS was glued by a two-component epoxy resin to the GFETs array area. The epoxy layer also protects wire bonds from unintentional damage during the chip operation.

*2.3 Assembling of the GFET aptasensor*

Prior to the assembling procedure, the graphene chips were treated for 5 min by low-pressure mercury lamp Svetolit-50 (power 50 W, LIT, Russia) to remove organic residuals (Emelianov et al., 2016). The PBASE molecule was used as a linker for the covalent binding of aptamers. For linker immobilization on graphene by π-π stacking, we used the transverse electrical field to increase the PBASE density (Hao et al., 2020). 4 mM PBASE solution in DMF was introduced into the PDMS well to soak graphene channels for 3 h at room temperature. Graphene channels of the GFETs were grounded while Pt wire under negative voltage with amplitude 0.3 V was inserted in DMF above the GFET array. The well was then rinsed consequently with DMF, IPA, and deionized (DI) water to remove reagent excess. 100 nM of anti-OTA aptamer dissolved in PBS solution (with pH = 7.4) was introduced into the well, and the chip on PCB was kept overnight (12 hours) in humid atmosphere to ensure aptamer binding to the PBASE linker. After rinsing several times in PBS solution to remove non-bound aptamers, 100 mM ethanolamine solution in PBS was introduced and kept for an hour into the well to block and deactivate non-bonded reactive groups.

*2.4 Aptasensor characterization and OTA detection*

Optical characterization was made on optical profilometer Huvitz HRM300 (Taiwan). The presence of adsorbed PBASE on graphene surface was investigated by microRaman spectroscopy (Centaur HR, Nanoscan Technology, Russia) with a 100x objective (NA=0.9) at a 532-nm wavelength (Cobolt, Solna, Sweden) with a beam spot of 1 µm$^2$ and laser power of 0.5 mW.

For liquid gate measurements, we introduced the Ag/AgCl pellet electrode (Science Products GmbH) with a diameter of 1 mm in PDMS well with the analyte. For transfer curves, the drain-source voltage ($V_{ds}$) was set to 20 mV, and the gate voltage ($V_g$) was swept with 10 mV resolution. Small sweep steps provide more accurate data points. During time-series measurements, the $V_{ds}$ is set to 20 mV, and $V_g$ is kept at 0 V until the drain-source current ($I_{ds}$) was stabilized. The real-time measurements started with 50 µL of blank PBS solution in a well. After GFET signal saturation, a drop of 30 µL of buffer solution was taken out, and the same volume of buffer-spiked OTA solution was added promptly to prevent channel drying. The latter steps of 30 µL solution exchange were repeated for the stepwise rising concentration of OTA. A simple drop-method of the sample analysis is more realistic, simplistic, straightforward, and can be nicely manufactured to fit in-field application package. The sensor's sensitivity was determined as $(R_0-R)/R_0=\Delta R/R_0$, where R is the resistance of the GFET channel and $R_0$ – the resistance of the GFET channel in an OTA-free solution.

Wine samples, bought in a local store, were diluted with 1x PBS in a 1:10 ratio without any specific pre-treatment. Two types of wine were investigated: dry red wine (pH=3.3) and semi-sweet white wine (pH<2.9). To prepare spiked wine solution, the proper concentration of OTA in PBS was added to diluted wine. In this way, concentrations from 6 pM to 500 pM were obtained. The analysis was carried out in the same way as described above.

## 3. Results and Discussion

*3.1 Device characterization*

The scheme of the GFET structure and its integrated design are shown in Fig. 1. In this GFET sensor, graphene works as a channel modified by PBASE-linked aptamers. We used the recently suggested method for PBASE controlled deposition under the applied transverse electric field to increase the PBASE coverage on graphene (Hao et al., 2020).

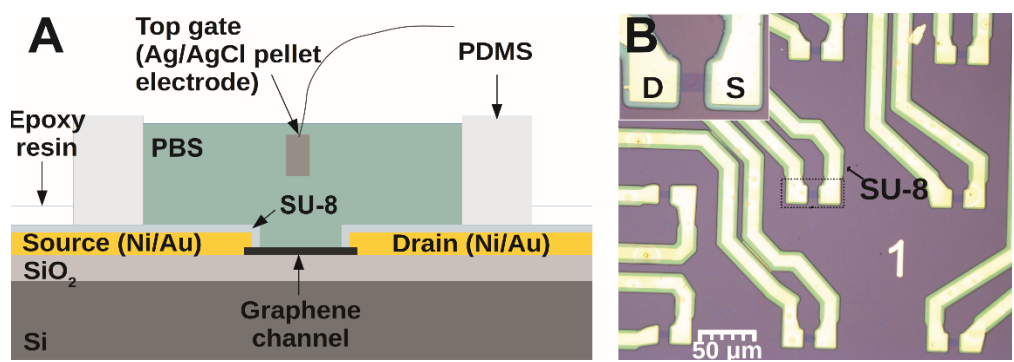

**Fig. 1.** Design of the sensor based on GFET. (**A**) A simplified scheme of a single liquid-gated GFET sensor cross-section. (**B**) Optical micrograph of the GFETs array on a chip with drain (D) and source (S) electrodes passivated by SU-8 (inset: the enlarged part of graphene channel).

Fig. 2A demonstrates the final assembled device with a GFET chip on PCB board. The electrical characteristics of the graphene FET after PBASE deposition and aptamer covalent binding are presented in Fig. 2B. After PBASE deposition, we observe the right shift of the Dirac point at about 75 mV due to p-type doping of graphene by a pyrene group (Wu et al., 2017). After aptamer covalent binding to PBASE, the G-rich backbone of oligonucleotides brings the negative charge close to the graphene surface, causing weak n-type doping of graphene (Danielson et al., 2020) resulting in the left shift of the Dirac point at about 40 mV. Overall decrease in the current at each assembly step can be associated with decreasing the number of the main charge carriers and increasing the number of the scattering centers. The latter is also causing the slope declination for both n-type and p-type branches on the transfer curves. The microRaman characterization brings more detailed information on nonlinear optical graphene properties change. There is a noticeable difference in Raman characteristics of graphene after conventional PBASE deposition and under an applied electrical field. The increase of the D band (1350 cm$^{-1}$) and appearance of the pyrene-related band (1628 cm$^{-1}$) (Liu et al., 2014) prove the PBASE deposition in both regimes. Relative intensity of this pyrene-assigned band to the G band can be used to quantify pyrene groups anchored on the surface. The preserving G band and the width of the 2D band confirm no significant damage of graphene that could be caused by chemical adsorption (Fig. 2C). Noticeable increase in the D band with preserving 2D/G ratio (Fig. 2D) show the rise in the coverage area of PBASE when deposited under the electrical field (Hao et al., 2020).

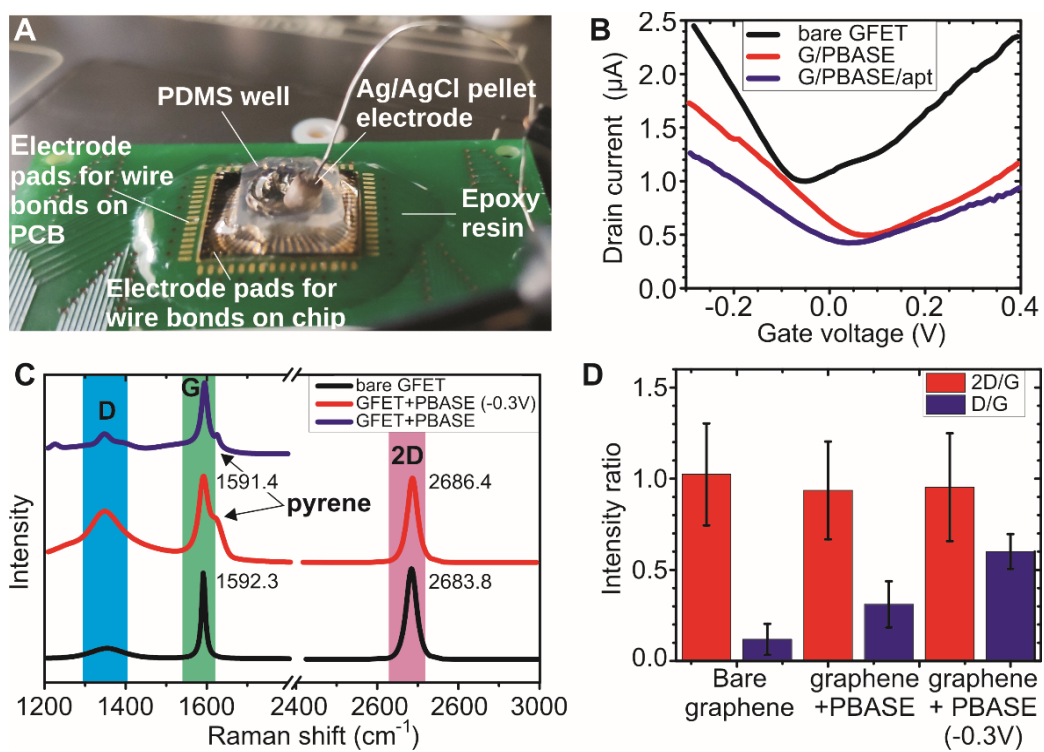

**Fig. 2**. Assembling the GFET aptasensor. (**A**) Photograph of the assembled sensor chip, wire-bonded on PCB with micro-well and Ag/AgCl pellet electrode. (**B**) Transfer curves of GFET at a drain voltage 20 mV for a bare GFET (black), after PBASE deposition (red), and covalent binding of aptamer (blue). (**C**) Raman spectra of bare graphene channel (black) and after PBASE deposition under -0.3V voltage applied (red) or without electrical field (blue). (**D**) The change in intensity of Raman peaks for bare graphene and after deposition of PBASE. Error bars are ±SD and result from multiple (N=3) measurements from different graphene areas.

The main mechanism of detection for the GFET-based aptasensor is the modulation of the channel conductance during the reconfiguration of the aptamer upon conjugation with small molecules. The thickness (Debye length, $\lambda_D$) of the electrical double layer (EDL) is very critical for the detection of small molecules, limiting the non-screened field-effect on the conductivity of the channel (Nakatsuka et al., 2018). The EDL thickness is highly dependent on the ionic strength (*I*) of the buffer and, for example, has a value of only 0.75 nm for 1x PBS (Nakatsuka et al., 2018; Sohn et al., 2013). The aptamer reconfiguration is an important process in sensor performance because it leads to the change in distance between electrical charge in molecules and graphene. Initial configuration of an aptamer on graphene surface is still in debate suggesting either free-standing position of the aptamer (Chen et al., 2019; Park et al., 2021) or π-π stacking of the backbone and aromatic bases of nucleic acids onto hexagonal carbon lattice (Béraud et al., 2021; Wang et al., 2012). Noticeable n-type doping (Fig. 2B) of graphene after aptamer deposition confirms the π-π stacking of oligonucleotides on graphene and direct electron transfer. To provide deeper insight into the mechanism of binding of the aptamer and small molecules, we performed experiments on MT detection.

*3.2 Mycotoxin detection*

In this work, we leveraged liquid-gate measurements of N=5 GFETs modified with anti-OTA aptamers for multimodal and rapid OTA detection. Electrical characteristics were obtained by drop-by-drop application of buffer solution with increased OTA concentrations, and using a classical Ag/AgCl pellet as a reference electrode. The regeneration of sensors was performed after all concentrations were measured. The transfer curves were recorded for OTA solutions in PBS with different ionic strength, *i.e.,* 1x PBS buffer (Fig. 3A) and 0.1x PBS (Fig. 3B). We use multiple measurements of I-V sweeps to stabilize the EDL before OTA insertion (Fu et al., 2017). Both the transconductance increase in the hole-doped region and the Dirac point shift were observed as a response to OTA concentration (Fig. 3C). The direction of the shift confirms the p-type doping of graphene due to reconfiguration of aptamers and decreasing the electrostatic effect on charge carriers. Shift of the Dirac point, increase of transconductance for p-type branch, and a slight increase of the minimal current suggest both the change in mobility and main charge carriers concentration in the graphene channel (Béraud et al., 2021). The electrostatic doping effect becomes higher when performing experiments in 10 times diluted (0.1x) PBS (Fig. 3B), that gives up to 10 folds increase of sensitivity for low concentrations. Interestingly, the transconductance has shown a nonlinear dependence on OTA concentration in 1x PBS, with current decreasing for concentrations above 100 pM. In a solution with high ionic strength, the charges from the G-quadruplex of the aptamer can be shielded from the graphene channel by the buffer ions because of a small Debye length due to high ion concentration (Béraud et al., 2021). However, it is still possible for the mycotoxin molecules to penetrate the EDL, dope the graphene non-specifically by direct adsorption and induce the decrease of mobility for both types of charge carriers. Increase in Debye length by dilution of the solution can significantly increase the efficiency of electrostatic gating from the aptamer, yielding the sensor's performance (Fig. 3D). The sensitivity to the OTA molecules with concentrations below 1 pM can be achieved in such a configuration (Fig. 3C). However, a small ionic strength of the buffer can limit the overall GFET-sensor application for MT detection in real samples. The dilution of a buffer acts unfavorably on aptamer unfolding as well as distancing from real sample environment (complex and high ionic concentration in wine, for example). Thus, real-time experiments were conducted in 1xPBS that is close to the ionic strength of wines (Yan et al., 2017).

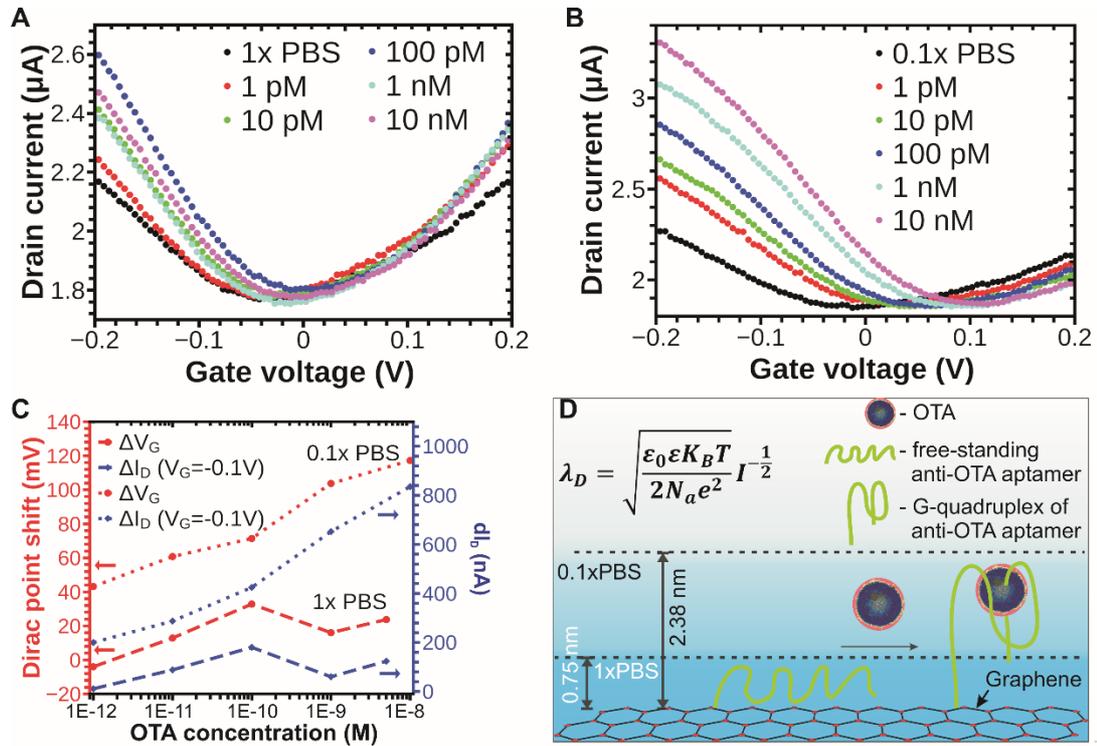

**Fig. 3**. Electrical characterization of the OTA sensor. (**A**, **B**) Transfer curves for different OTA concentrations in 1x PBS (**A**) and 0.1x PBS (**B**), both with $V_{ds}$=100 mV. (**C**) Change in the Dirac point and drain current (at $V_G$=-0.1V) for increased OTA concentrations in buffer solution with different ionic strength. (**D**) Hypothesized mechanism of anti-OTA aptamer target-induced reconfiguration close to graphene channels with different ionic strength. Aptamers π-stacked on graphene form G-quadruplex in the presence of OTA detracting from graphene channels, thereby increasing the transconductance and the Dirac point shift.

We studied the real-time response of the aptamer-modified GFETs while adding the OTA solutions with different concentrations. The current change in time has been recorded in parallel from N=5 similar GFETs (Supporting Information: Fig. S1), and the statistically relevant data is presented in this work. As shown in Fig. 4a, there is a noticeable and rapid response for small concentrations of OTA. The response is saturated for concentrations above 100 pM (Fig. 4B). Our experiments show that after 80 pM of added OTA, majority of the available receptors are saturated due to the small graphene channel size and the limited number of aptamers. The effect is reproducible and supported by the Dirac point shift saturation at 100 pM derived from $I_D$-$I_G$ sweeps, especially when 1xPBS is used, as shown in Figure 3C. Variations of the height of EDL layer may also affect on the electrostatic shielding resulting in a different level of saturation. Similar behaviour is widely reported in the literature (Chen et al., 2019; Danielson et al., 2020; Nakatsuka et al., 2018). The current increase for the aptamer-modified GFET upon OTA binding is noticeable when working in a p-doped regime with the Dirac point right shift as shown in Fig. 3A,B. To provide the control measurements, we performed similar experiments with bare non-functionalized GFET (Supporting Information, Fig. S2), and clearly no response to OTA is visible. The weak resistance change for concentration higher than 4 nM is due to OTA adsorption on graphene surface where phenolic moiety of the toxin can undergo π-stacking interaction (Fadock and Manderville, 2017) and results in direct doping of graphene. The results demonstrate that the sensor can work both following the Dirac point and drain current changes. The drain current real-time monitoring provides a simple electronic tool for rapid and accurate measurements of OTA concentration in in-field experiments.

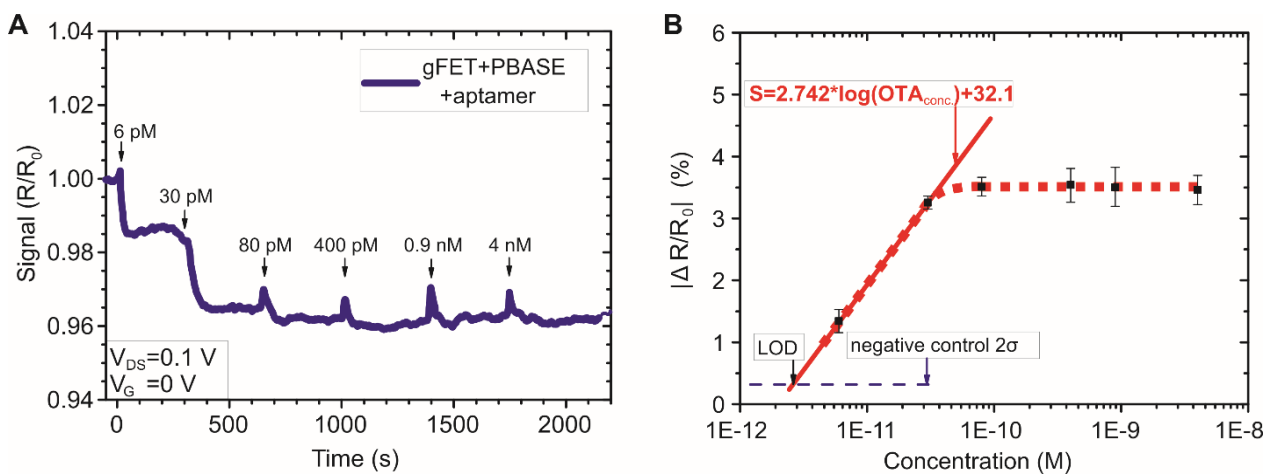

**Fig. 4**. Real-time measurements of OTA solutions in 1x PBS. (**A**) Time course of GFET sensor response to increasing OTA concentration. (**B**) The calibration curve for sensitivity to OTA is based on the data from GFET, shown in (A).

In general, two mechanistic concepts of GFET sensing are discussed that may take place due to aptamer conformational changes. Upon target capture, aptamer three-dimensional structural reorientation occurs in such a way that substantial portions of the negatively charged backbones are moved either closer to or further away from the sensing channels (Nakatsuka et al., 2018). In the case of the anti-OTA aptamer, the mechanism of sensing is based on a reconfiguration of the aptamer to G-quadruplex upon binding to OTA (Fadock and Manderville, 2017; Wang et al., 2019), which is withdrawn from graphene and increase the distance from the charged aptamer and the surface, as shown in Fig. 3D. This effect can be associated with the small size of the OTA molecule (403.82 Da) compared to the longer aptamer molecules. Electrostatic gating effect in the graphene channel is rapid, yielding a short response time.

At low concentrations (1 – 100 pM), the aptamer first captures the OTA molecules due to their strong affinity. In response, the GFET's charge neutrality point shifts to the more positive values with OTA concentration increasing, indicating the hole doping upon the aptamer folding. Note that the maximum value of the Dirac shift for OTA in 1x PBS (Fig. 3C) is about 30 mV that is close to the initial shift due to the aptamer covalently binding to PBASE during sensor assembly. The findings support our suggested hypothesis of the partial withdrawal of the aptamer from graphene surface during the unfolding of the adsorbed state to the G-quadruplex. Some effects from the OTA charge can be noticeable but only for large concentrations of OTA above 1 nM. For higher concentrations, the impact of mycotoxin is based not only on the electrostatic modulation of the charge carriers in the channel by the aptamer reconfiguration but also on the accumulated mass/charge effect of MT molecules on aptamer-free graphene.

*3.5. Analytical performance of the biosensor*

Parallel experiments were carried out on an array of N=5 identical GFET sensors on a single chip. The sensor's relative response versus OTA concentration is plotted in Fig. S1. All five devices displayed very similar properties and biosensing response, indicating excellent reproducibility. We measured the current amplitude variation of devices in the range of 6 pM to 7 nM with the linear relation to the target concentration in the range of 6 - 100 pM in 1x PBS. The noise level was calculated as relative standard deviation (SD) of the background signal of the time course curve for negative control and equaled to 0.1%. The analytical LOD was determined as the intersection of the sensitivity in the linear regime and two standard deviations (2σ) of the signal of negative control (Orlov et al., 2017), as shown in Fig. 4B. The calculated LOD is 1.4 pM. This detection limit is far below the tolerance level of OTA in food that is about 5 nM (Gil-Serna et al., 2018). For calculation of response time to OTA, we used the single exponential fitting for curves presented in Fig. 5A. The time of response of our GFETs to the OTA varies from 10±3 s for 6 pM to 37±13 s for 7 nM (Fig. 5B). The slowing down in the speed of change in the drain current for higher concentrations can be associated with the saturation of all receptors already for 100 pM of OTA due to the small dimensions of graphene channels. The slight variation of time for different concentrations can be associated with liquid kinetics when replacing the solution, especially when most OTA binding sites are occupied (at high MT concentration). This effect shows high reproducibility on the array of measured aptasensors on a single chip (see Fig. S1). However, the measured response time does not specifically scale with OTA concertation (see Fig. 5B), and on average, the value is <60 s is still superior to the most existing mycotoxin detection methods (300-1800 seconds, see Table 1). The GFET aptasensor performance in terms of speed and LOD is superior to previously reported aptasensors based on electrochemical or fluorescence methods (see comparative Table 1).

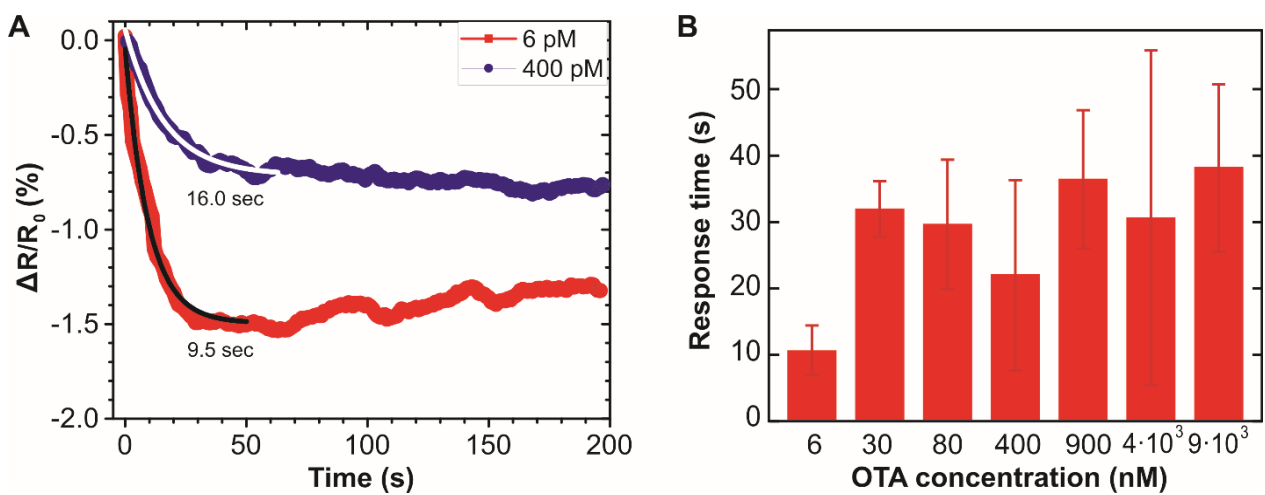

**Fig. 5.** Time course of resistance change of GFETs modified with the OTA aptamer. (**A**) Experimental data of response for 6 pM (red) and 400 pM (blue) and the single exponential fitting curves (black). (**B**) The response time of aptasensors array of five GFETs for different OTA concentrations.

Extraordinary sensitivity of the GFET aptasensors can be explained by the increase of the binding sites for the aptamers during the electric field-supported deposition of PBASE (Hao et al., 2020). An unprecedented sensitivity of the device is due to the high response of the graphene channel to the smallest variation of electrical charges in the surface's vicinity. Also, the antiparallel G-quadruplex aptamer topology upon OTA binding provides a more pronounced reconfiguration of the aptamer, effectively modulating the conductivity in the GFET channel. The suggested assay outperforms the recently reported approaches, making it a promising platform for proper aptamer selection and real-time kinetics investigation (dos Santos et al., 2019).

**Table 1**. Comparison of the OTA detection in red wine for different reported aptasensors

| Detection Method | Assay with aptamer | Red Wine Sample Preparation | Response Time, s | Limit of Detection, pM | Dynamic range | Ref. |
|---|---|---|---|---|---|---|
| FET | Graphene/PBASE | Centrifugation | 300 | 10 | 10 pM – 10 nM | (Nekrasov et al., 2019) |
| fluorescence | ZnPPIX probe | Methanol extraction | - | 30 | - | (Liu et al., 2018) |
| TIRE | Glass/gold | - | 600 | 30 | - | (Al Rubaye et al., 2018) |
| CL | Signal probe/hemin | $H_2O_2$/KCl | - | 200 | 0.25 – 5 nM | (Wang et al., 2019) |
| EC | CdTe\graphene\Au | - | 1800 | 0.2 | 0.5 pM – 10 nM | (Hao et al., 2016) |
| EC | Au NPs\methylene blue | Filtration | 1800 | 0.75 | *2.5 pM–2.5 nM* | (Yang et al., 2014) |
| EC | Gold-dsDNA/g-CNNS | Filtration | 1800 | 73 | 0.2-500 nM | (Zhu et al., 2018) |
| EC | GCE-carboxylated G-Janus particles | Dissolved in buffer drop | 1800 | 0.01 | 10 fM – 10 nM | (Yang et al., 2019) |
| FET | Graphene/PBASE | No preparation | 10 | 1.4 | 5-500 pM | *This work* |

CL-chemiluminescence. EC-electrochemical. ZnPPIX–Protoporphyrin IX Zinc (II). NPs - nanoparticles. GCE-glassy carbon electrode. CNNS - graphitic carbon nitride nanosheet. dsDNA – double-stranded DNA

The increased speed of detection and the lower detection limit can be associated with the analyte solution injection as well. During experiments, the speed of a jet flow out of the micropipette channel can be two orders of magnitude higher than the quasi-laminar flow in the pipette, reaching several meters per second (Priebe et al., 2010). We hypothesize that the high degree of turbulence present during solution injection may influence the effective mixing of the aptamer molecules. Since the aptamer molecules possess high affinity, the effective mixing may lead to rapid capturing of OTA, even in the highly diluted solutions, due to the disturbance of the dynamic balance in small molecules distribution in the solution.

*3.6. Selectivity and Specificity*

The selectivity of the sensors was studied on AFM1 mycotoxin as a nonspecific target for the OTA aptamer. The AFM1 was introduced in the sensing area through the same dynamic experimental process used for the OTA assay (Supporting Information, Fig. S3). No response was detected for the concentrations of AFM1 in the range from 1 pM to 1 nM, with only a weak response detected for AFM1 concentration higher than 1 nM, which can be explained by direct nonspecific charge transfer from MT onto graphene. Thus, the method demonstrates acceptable specificity of the sensor.

Besides, we have investigated the effect of cross selectivity to OTA. The two 100 pM OTA solutions were prepared in pure PBS and PBS additional with spiked 100 pM of AFM1, respectively. The response to the two buffer solutions was estimated as 2.4±0.4% and 2.8±0.5% for OTA and OTA+AFM1 solutions, respectively. Thus, the addition of interfering toxins does not alter the sensor signal due to the high affinity of aptamers, confirming high device selectivity.

*3.7. Real samples results*

Finally, to demonstrate the performance of developed sensors for the real product analysis, we spiked wine samples with mycotoxins. The high complexity of the wine matrix that consists of different chemicals both of the non-volatile nature (polyphenolic molecules such as tannins, pigments, polysaccharides, proteins, alcohols, acids) and volatile aromas makes it analytically challenging samples (Yan et al., 2017). In our trials, we suggest an easy-to-use method that requires wine to be simply dissolved in a buffer. The best agreement with the calibration curve (90 ± 10%) was demonstrated for a low concentration of OTA, 6 pM (Fig. 6A). The response times were estimated as 50 s for 6 pM and 74 s for 60 pM (Fig. S4) that are longer compared to the PBS experiments, suggesting the effect of the complex matrix of the wine limits the aptamer-OTA binding reaction velocity. Complex ion composition of wine can also affect the EDL formation near the graphene surface (Lee et al., 2015). We observed the weak degradation of GFETs characteristics even with washing steps during the real-time measurements. Nevertheless, for proper restoration of the transistor after measurement, we applied several washing cycles in urea, PBS, and DI water because of the complex wine structure. This limitation on the sensor performance can be overcome by proper pH and temperature control during sensing (Fakih et al., 2020). Importantly, the reported 6 pM OTA detection in 10% wine solution translates to 60 pM of OTA detection in undiluted wine, which is still two orders of magnitude lower the EU legal tolerate level for OTA presence in wine (Gil-Serna et al., 2018).

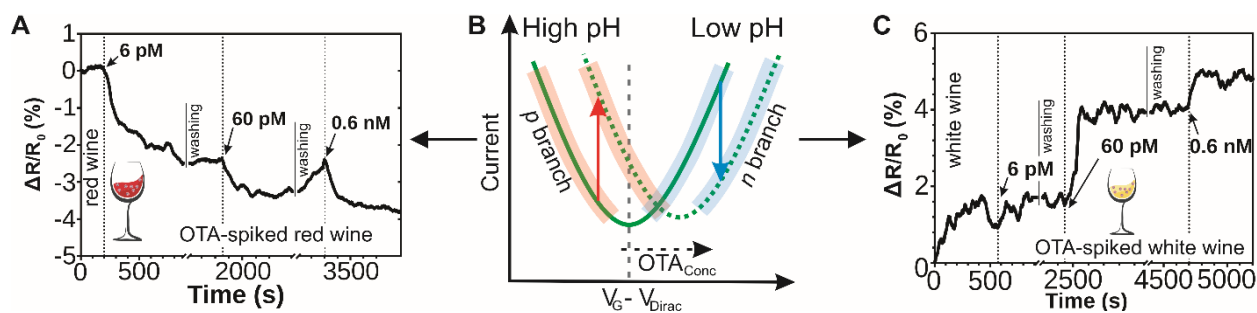

**Fig. 6.** Real-time detection of the increased concentration of OTA in red wine (A) and white wine (C) with the anti-OTA aptamer modified GFET. (B) The scheme of low and high pH effects on GFET biosensor response.

Interestingly, the signal was opposite for the red and white wine samples spiked with OTA (Fig. 6C). We assume that low pH of white wine (less than 3) can shift the Dirac point to the left, resulting in changing the type of main charge carriers (Lee et al., 2015; Sohn et al., 2013). When graphene becomes *n*-doped at low pH, the reconfiguration of G-quadruplex of aptamer increases the holes in the channel, thus the sensor's resistance grows (Fig. 6B). The electron mobility in our GFETs is lower compared to hole mobility; hence the sensor sensitivity at the side is lower. The complex matrix of wine could potentially cause the degradation of the sensor response due to the organic residuals even after the washing steps. To corroborate the findings, we performed the experiments in OTA spiked wine with bare GFET without any immobilized aptamer. As demonstrated in Fig. S5, there is no observable response; hence the results shown in Fig. 6 are a specific response of the functionalized graphene with OTA. In overall, the developed devices provide the LOD of OTA in wine as low as 1 pM and 3 pM for red and white wine, respectively. The proper consideration of complex analyte composition is the calibration of a sensor that takes into account the properties of graphene and the solution. The best option is to use an array of sensors, where one GFET can be a reference providing real-time differential analysis. Moreover, the suggested approach for sensor development is scalable and can be used to produce low-cost disposable chips for Point-of-Care technologies for food safety analysis.

## 4. Conclusions

A novel aptasensor based on graphene FET array for mycotoxin detection has been demonstrated. It provides an accurate and rapid electrical response when the aptamer binds to the target small molecules changing its secondary configuration in the process. The challenging liquid-gate configuration in terms of liquid properties, such as pH and ionic strength, was evaluated as a crucial for GFET performance. Ionic strength modulation can affect the sensor's performance positively regarding the sensitivity and credibility of the sensor for real-sample investigation. The response time down to ~10 s was demonstrated in the phosphate buffer, increasing up to ~50 s for real life samples such as wine. The described device demonstrates unprecedented sensitivity to OTA, with a LOD as low as 1.4 pM in PBS and 1 pM in wine. The negligible cross-reactivity of OTA aptasensors to a different mycotoxin, like AFM1, makes devices suitable for food and feed analysis. The straightforward sample droplet method greatly simplifies the assay preparation even for liquids with a complex matrix. The suggested approach based on the array of GFETs could be further developed to detect multiple targets on a single chip after proper patterning and assembling with several mycotoxin-specific aptamers as well as to conduct differential analysis that will not demand prior sensor calibration.

## Acknowledgments

This work was supported in part by the IPANEMA project, which received funding from the European Union's Horizon 2020 research and innovation programme under grant agreement N° 872662, in part by the Russian Science Foundation under grant numbers 19-19-00401 (graphene chip development and characterization), and 21-12-00407 (monitoring analyte-aptamer interactions and investigation of aptasensor parameters). I.G. and I.B. participated in a project that has received funding from the European Union's Horizon 2020 research and innovation programme under grant agreement N° 739570 (ANTARES). S.J. acknowledges the financial support of the Ministry of Education, Science and Technological Development of the Republic of Serbia (Grant Nº. 451-03-9/2021-14/200358).

# Supporting Information

# Real-time detection of Ochratoxin A in wine through insight of aptamer conformation in conjunction with graphene field-effect transistor

**Nikita Nekrasov**[a#]**, Stefan Jaric**[b#*]**, Dmitry Kireev**[c]**, Aleksei V. Emelianov**[a]**, Alexey V. Orlov**[d]**, Ivana Gadjanski**[b]**, Petr I. Nikitin**[d]**, Deji Akinwande**[c] **and Ivan Bobrinetskiy**[a,b*]

[a]National Research University of Electronic Technology, Moscow, Zelenograd, 124498, Russia, 8141147@gmail.com

[b]BioSense Institute - Research and Development Institute for Information Technologies in Biosystems, University of Novi Sad, Novi Sad, 21000, Serbia

[c]Department of Electrical and Computer Engineering, The University of Texas at Austin, Austin, TX, USA

[d]Prokhorov General Physics Institute of the Russian Academy of Sciences, 119991, Moscow, Russia; petr.nikitin@nsc.gpi.ru

*Corresponding author: e-*mail*: bobrinet@gmail.com; e-*mail*: sjaric@biosense.rs

# these authors contributed equally

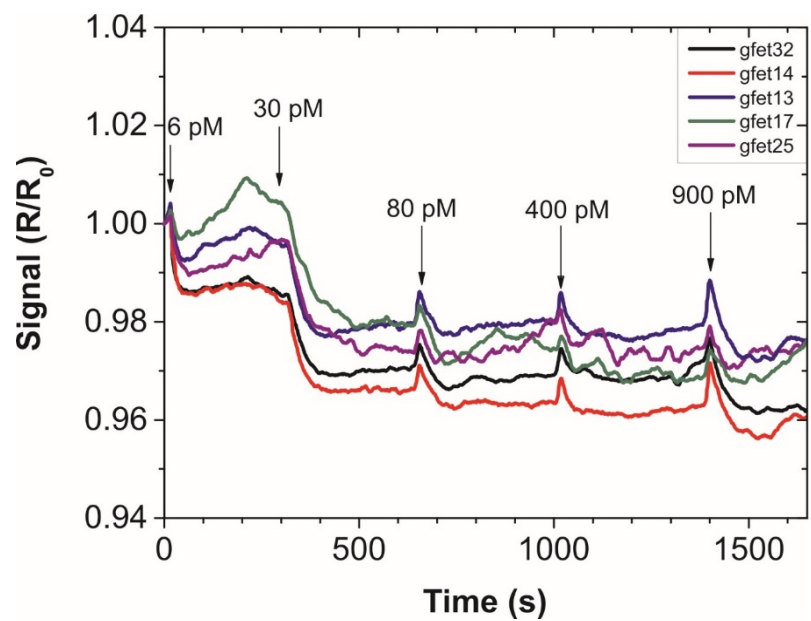

**Fig. S1.** Time course of response of an array of five GFET sensors under increasing OTA concentration in 1xPBS.

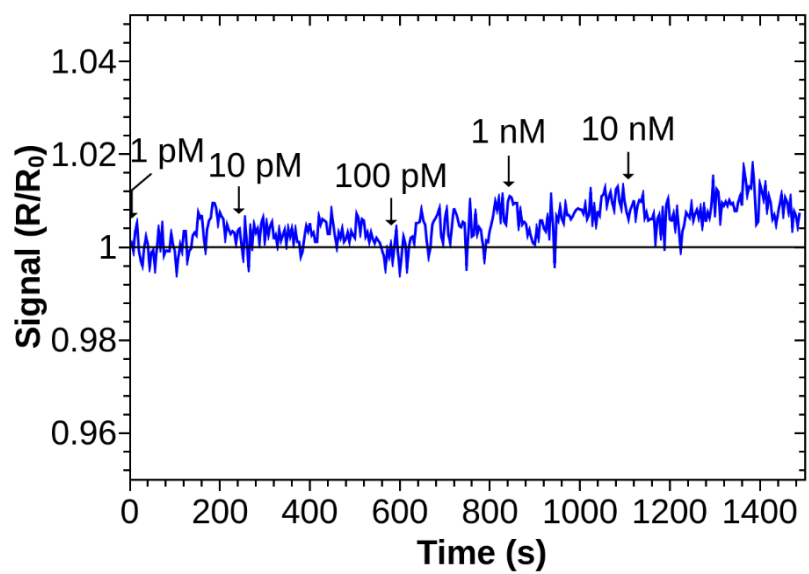

**Fig. S2.** Time course for bare (non-functionalized) GFET for different OTA concentrations in 1xPBS.

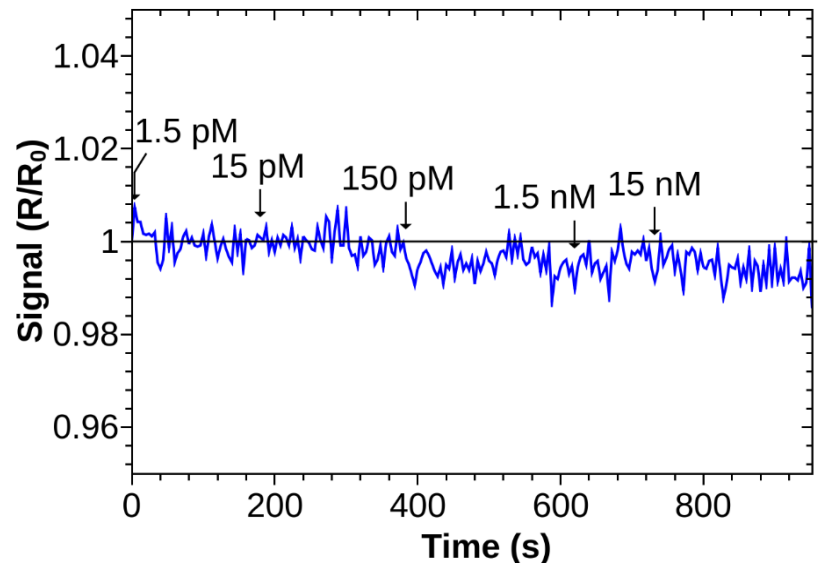

**Fig. S3.** Control experiments. Response time course for anti-OTA aptamer-modified GFET when different concentrations of AFM1 in 1x PBS are applied.

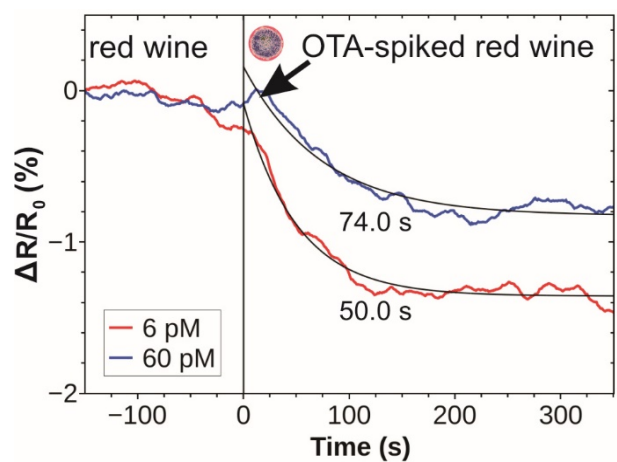

**Fig. S4**. Time course curves of GFET sensor response to red wine samples spiked with OTA and its exponential extrapolation for response time analysis. The zero time corresponds to the insertion of spiked wine.

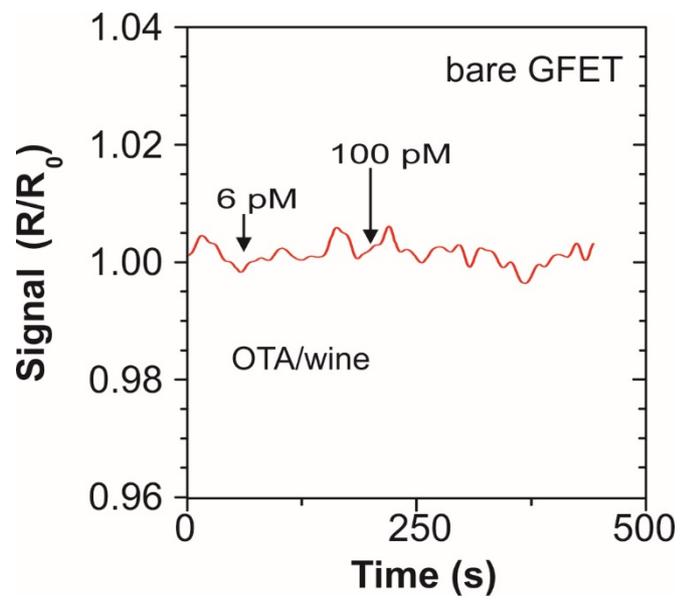

**Fig. S5**. Time-course of bare GFET in red wine with increased OTA concentration